\documentclass{article}
\usepackage{arxiv}

\usepackage[utf8]{inputenc} 
\usepackage[T1]{fontenc}    
\usepackage{hyperref}       
\usepackage{url}            
\usepackage{booktabs}       
\usepackage{amsfonts}       
\usepackage{nicefrac}       
\usepackage{microtype}      
\usepackage{lipsum}
\usepackage{graphicx}
\usepackage{footmisc}
\graphicspath{ {./images/} }

\title{Online Hackathons as an Engaging Tool to Promote Group Work in Emergency Remote Learning}

\author{
 Kiev Gama, Carlos Zimmerle, and Pedro Rossi \\
  Centro de Informática\\
  Universidade Federal de Pernambuco (CIn/UFPE)\\
  Recife, Brazil \\
  \texttt{\{kiev,cezl,pgrr\}@cin.ufpe.br} \\
   \\
}

\begin{document}

\maketitle
\begin{abstract}
 In 2020, due to the COVID-19 pandemic, educational activities had to be done remotely as a way to avoid the spread of the disease. What happened was not exactly a shift to an online learning model but a transition to a new approach called Emergency Remote Teaching. It is a temporary strategy to keep activities going on until it is safe again to return to the physical facilities of universities. This new setting became a challenge to both teachers and students. The lack of interaction and classroom socialization became obstacles for students to continue engaged. 
 Before the pandemic, hackathons -- short-lived events (1 to 3 days) where participants intensively collaboration to develop software prototypes -- were starting to be explored as an alternative venue to engage students in acquiring and practicing technical skills. In this paper, we present an experience report on the usage of an online hackathon as a resource to engage students in the development of their semester project in a distributed applications course during this emergency remote teaching period. We describe details of the intervention and present an analysis of the students' perspective of the approach. One of the important findings was the efficient usage of the Discord communication tool -- already used by all students while playing games -- which helped them socialize and keep them continuously engaged in synchronous group work,``virtually collocated''.
  
\end{abstract}


\section{Introduction}

As a consequence of the COVID-19 pandemic, in the year of 2020, educational institutions worldwide canceled their face-to-face classes and moved to online courses to avoid the spread of the disease. Online learning is an educational approach that has been conceived with a different purpose and specialists are rather calling this new model as \textit{Emergency Remote Teaching} -- or learning, from the students perspective --  which involves the use of remote teaching solutions for instructions that would otherwise be delivered face-to-face if there was no pandemic
~\cite{hodges2020difference}.

Many recent reports about this new model are sharing common findings of students low motivation and engagement in emergency online teaching. There are many challenges to students with this sudden change: technological constraints, lack of a sense of belonging and connectedness, presence of distractions at home, lack of engagement \cite{xie2020covid}. The approach has drawbacks in social aspects such as the lack of interaction with the instructor and the absence of classroom socialization \cite{muhammad2020covid}. Those are all fundamental aspects since learning is a social and cognitive process \cite{hodges2020difference}. Aspects like motivation, self-efficacy, and cognitive engagement decreased after the transition to emergency online learning, which demands educators to be mindful of these new circumstances and promote a positive attitude, encouraging students' motivation \cite{aguilera2020college}. Being aware of that, we decided to experiment an approach to stimulate more intense online social interactions among students in the context of an elective (i.e., optional) course, by utilizing a hackathon format for the development of their project.

Hackathons are short-lived events which take typically 1-3 days \cite{komssi2015hackathons}, frequently happening on weekends, where teams develop functioning software prototypes tailored to specific challenges. Due to its practical and challenge-oriented nature, it has attracted computing-related undergraduate students as an informal venue for learning~\cite{nandi2016hackathons}  and more recently being adopted in undergraduate courses \cite{uys2019hackathons,gama2018hackathons,porras2018hackathons}. One of the main characteristics of hackathons is the radical collocation of participants~\cite{trainer2016hackathon}, which is something that would have to be adapted in an online scenario.

This paper describes our experience in conducting an online hackathon with 18 students for their semester project in Distributed Applications course. We used a mixed-method approach to gather and analyze feedback from students about their perception of this intervention. One of the most important findings was the efficient communication using the Discord tool, which was originally conceived for the gaming community. The voice channels and screen sharing features allowed to create a sense of virtual collocation (i.e., online)~\cite{olson2000distance}  where students could socialize, work synchronously, solve problems and learn together and even do pair programming~\cite{williams2001support}. 

\section{Background and Related Work}
Hackathons have been used as alternative venues for students practicing or acquiring new skills \cite{nandi2016hackathons, fowler2016informal}. These events started to be reported as being used by teachers as part of official class activities \cite{uys2019hackathons,gama2018hackathons,porras2018hackathons,horton2018project}. As an educational tool, we understand it with the potential to exercise the four abilities of experiential learning defined by Kolb \cite{kolb2014experiential}. It can naturally exercise concrete experience and active experimentation abilities and if appropriately guided, can explore reflective observation and abstract conceptualization. 

Many online hackathons are targeting problems related to COVID-19~\cite{vermicelli2020can} --  some of these events focusing on universities ~\cite{bolton2020virtual,hossain2020accelerating} -- but literature is lacking reports on such format as part of undergraduate course projects during the pandemics. So far, a framework for online hackathons was proposed in that scenario, but it lacks details or any concrete usage in a classroom or course setting~\cite{goodman2020learn}.

Radical collocation of participants and face-to-face interactions are key characteristics that help participants quickly advance technical work in hackathons \cite{trainer2016hackathon}. Replicating these dynamics in an online setting is challenging: physical collocation and face-to-face collaboration are not possible. Anyhow, many online hackathons are being held to crowdsource solutions to problems around COVID-19~\cite{vermicelli2020can}. Software development teams working remotely can be traced back to the idea of Global Software Development (GSD), where communication is one of the main challenges~\cite{herbsleb2001global}. Recent improvements in communication tools, such as Slack (\url{slack.com}), brought benefits to teams doing GSD \cite{stray2019slack}.

Slack is also being experimented with as a support communication tool in face-to-face classes, but it is not the most popular one among students \cite{menzies2020professional}. Other tools such as Discord (\url{discord.com}), originally conceived for game playing, are more used by students. Discord has been successfully used as a team communication tool in Game Jams --game-focused development events that are similar to hackathons-- even in cross-border teams \cite{fowler2020jamming}. Before the pandemics, there were reported attempts of using Discord as a support communication tool in the context of face-to-face learning \cite{lacher2018using} and, during the pandemics it has been used in online hackathons~\cite{goodman2020learn} and was promoted to be used in remote classrooms~\cite{discordblog}.

\section{Course Structure}
This optional course was available to Computer Science and Information Systems students under the name of "Advanced Topics in Information Systems". It consists of an introduction to Distributed Applications, aiming to provide undergraduate students from 3rd and 4th year with an overview and practical background on concepts and technologies to build distributed applications. There were 22 enrolled students but only 18 of them finished the semester. 

The course is divided into synchronous and asynchronous communication principles and technologies. \autoref{tab:coursecontents} details the technical topics and content. The first three topics (Sockets, RPC and HTTP/REST)  cover synchronous communication while the remaining four (MOM, WebSockets, EventSource, and Reactive Programming) consist of asynchronous communication. All examples, exercises and the project were developed using JavaScript/Node.JS. Student grades were composed by practice through individual assignments and a group project. The assignments were all individual except for a group project at the end of the semester. The idea with the assignments was putting students in contact with different technologies that are used in industry (e.g., gRPC, RabbitMQ, RxJS).

 
\subsection{Classes and Assignments}

Google Classroom was used throughout the semester for communication between teacher and students, class materials upload, and assignments. For the pedagogical approach, we alternated synchronous classes (expository) and technical assignments (practical), 
with the semester being finalized with a group project (collaborative) conducted under an authentic hackathon format.

\begin{table}[!ht]
  \caption{Technical topics covered in the course.}
  \label{tab:coursecontents}
  \begin{tabular}{p{15cm}}
   \toprule
   \textbf{Topic/Content}\\
    \midrule
    \textbf{Sockets:} Overview of TCP and UDP sockets for interprocess communication and patterns for the creation of communication protocols. 
    \\
    
    \textbf{Remote Procedure Calls and gRPC:} Introduction to RPC and Interface Definition Language (IDL) with practice on gRPC and Protobuffers.
    \\
    \textbf{HTTP/REST:} Basics of request/response, methods, headers and status codes. REST architectural principles 
    and its relationship with HTTP.\\
    
    \textbf{Message-Oriented Middleware:} Notion of indirect communication (publish/subscribe, message queues), with exercises using RabbitMQ.\\
    
   \textbf{WebSockets:} Basics of the technology and its versatility to be used either on servers or browsers.
    \\
    
   \textbf{EventSource:} Introduction to the EventSource interface, the concept of server-sent events and its utility for creating real-time data feeds.
   \\
    
    \textbf{Reactive Programming}: Overview of the approach, with examples on RxJS, and its importance while dealing with event-driven applications. 
\end{tabular}
\end{table}

Every Monday consisted of a synchronous class over Google Meet. 
The following Thursday consisted of an individual practical assignment focused on Monday's topic, to be turned in within a week. Each assignment accompanied a small list (3 to 5 links) of guiding resources (e.g., third-party tutorials and videos on the topic). Each assignment had between 4 to 6 tasks. The first half of each assignment employed a scaffolding approach, giving step-by-step instructions. The remaining half was in the form of specific challenges stimulating students' autonomy and problem-solving skills, that could eventually involve topics previously seen (e.g., create a chat application that combines WebSockets and EventSource). The end of the assignment had brief questions to encourage students to reflect on what they were learning (e.g., asking the main differences between the strategy taken to solve the current challenge when compared to the one from the previous assignment). 
Based on the demand of students, Monday classes were also used to discuss the previous Thursday's assignment with the teacher and peers.

\subsection{Online Hackathon}
The course project was intended for students to apply in a practical scenario the concepts and techniques seen in the course. Instead of a typical asynchronous course project (e.g., 2 to 4 weeks to be completed), we proposed a hackathon to create an opportunity for students to intensively collaborate in a synchronous way with continuous support from a mentoring team. The event was structured based on the teacher's experience in hackathons organization. We used a 10h hack day (in-person) in a previous offering of this course~\cite{gama2019developing}, with support from the teacher without mentors. This time, the course hackathon was conducted as a 4-day (\textasciitilde{}96h) online event, from November 19 to 23 in 2020 (Thursday to Monday, which were the official weekdays of the course), with the support of students as mentors. The event started during official class hours, at 5 PM on a Thursday. We adapted a code of conduct\footnote{\url{https://hackcodeofconduct.org/}} and gave instructions to students as detailed next.

\noindent\textbf{Team formation.} 
Students had to work in groups of 3 or four members, as announces weeks before. In majority, teams were formed before the hackathon. The exception was two students who started the hackathon without a team and choose to form a duo. In the initial hour, they decided to split and join separate groups. During instructions, we enforced the importance of having a rotation of activities so everyone can contribute to each part of the project. 

\noindent\textbf{Challenge.} The challenge was creating a project motivated by a real-world scenario where a distributed application uses streaming data from one or more sources within a domain of their interest (e.g., IoT, Urban Mobility, Water, Energy). We provided students with a list containing dozens of data sets but they could choose other sources.
They had to develop an end-to-end infrastructure demonstrating the following: the mediation of streamed data; triggering of events based on simple analysis; displaying of these events, alerts and data synthesis in a User Interface (e.g., a dashboard). 

\noindent\textbf{Communication tool.}
The official communication tool for the hackathon was the Discord chat application. It allows the creation of voice channels, where participants can talk simultaneously and start live transmissions where they share their screens. We configured our Discord server with one private channel for the mentoring team and two public voice channels: \#lounge (other conversations or game playing); and \#auditorium (presentations). There were text channels for announcements (\#general), \#off-topic and channels for technology discussion (\#http-rest, \#websockets-sse, \#gRPC, \#rabbitmq, \#rxjs, \#nodejs, \#ui). Each team had one public voice channel (i.e., anyone could join) and one private text channel, only visible to team members. We urged them to use only Discord for communication and to avoid other tools (e.g., Telegram, Classroom).

\noindent\textbf{Mentoring and Checkpoints.}
The mentoring team was composed by the teacher and three students that have taken this course in previous semesters. The student mentors were mostly available late in the afternoon and during the evening hours.  There were checkpoints with the teacher on Friday (7 PM), Saturday (11 AM and 7 PM), Sunday (11 AM and 7 PM) and Monday (2 PM). The first checkpoint was public in the auditorium channel, aiming to make groups aware of each other's choices concerning domains and data sets. The other checkpoints were private sessions in each team's voice channel. In both mentoring and checkpoint sessions, the mentor would easily switch from channel to channel, join ongoing conversations and ask about status or difficulties.

\noindent\textbf{Final presentation.}
The presentation happened on a Monday (7 PM). It had a free format (e.g., a pitch) but needed to cover the following: (1) context of the problem; (2) explanation of the solution features; (3) solution architecture; (4) list of used technologies/concepts; (5) live demonstration (video as backup); (6) GitHub repository link. Grading took place in the days after the event, but there was a symbolic prize for the best solution. The mentors considering the following criteria in the evaluation: technical complexity of the solution (architecture, level of use of involved technologies); impact in people's lives; visual appeal; and presentation.

\section{Data Collection and Analysis}
\label{sec:data_analysis}
We used a mixed-method approach, with both quantitative and qualitative data analysis from multiple data sources. An anonymous survey sent to students collected quantitative and qualitative data, and two other sources gathered qualitative data: participant observation and a collective reflection done in a project \textit{postmortem} session. 
We separately analyzed qualitative and quantitative data and then converged results during interpretation \cite{creswell2007designing}.

\noindent\textbf{Survey.} We sent an anonymous questionnaire by e-mail the day after the hackathon presentation.  We avoided asking the respondent's gender because they were all men except for one non-binary whose identity would be exposed by such a question. The first part of the survey had a disclaimer, general questions about the experience in hackathons, and their opinion on using Discord. The second, third, and fourth parts asked about the hackathon format; students' motivations; and aspects about working remotely, respectively. In total, those parts contained 18 statements (\autoref{tab:questions}) on a Likert scale response format (Strongly Disagree, Somewhat Disagree, Neutral, Somewhat Agree, Strongly Agree). The fifth part of the survey focused on learning. Based on the revisited Bloom's Taxonomy~\cite{anderson2001taxonomy}, which categorizes six levels of thinking (remembering, understanding, applying, analyzing, evaluating, and creating) that learners may have about a given subject or topic, we asked students to rank their perception of learning on each technical topic (\autoref{tab:coursecontents}) on three moments: (1) before the course, (2) right before the hackathon, and (3) after the hackathon, so we could understand their perceived evolution. We (\autoref{fig:bloom}) adapted a questionnaire based on Bloom's taxonomy for student's self-assessment in programming~\cite{alaoutinen2010student}. 
Three open-ended questions asked about: the usage of Discord, the positive, and the negative aspects of the hackathon. Since answers were short sentences, two researchers did a simple thematic analysis to identify similar codes on students' answers and to generate themes.

\begin{table}[!ht]
  \caption{Questions using a Likert scale response format. }
  \label{tab:questions}
  \begin{tabular}{lp{15cm}}
    \toprule
    \textbf{ID} & \textbf{Topic/Question}\\
    \midrule
    \multicolumn{2}{l}{\textbf{Motivation}}\\
    Q1  & I was excited to participate in the hackathon\\
    Q2  & During the hackathon, I felt motivated (motivation being more personal, more intrinsic, linked to your interest)\\
    Q3  & During the hackathon I felt engaged (engagement being more visible, as your involvement and commitment to the group's goals)\\
    \multicolumn{2}{l}{\textbf{Hackathon as a project format}}\\
    
    Q4  & From my experience (if you've participated in hackathons) or from what I've heard (if you haven't participated), this hackathon was truly authentic, resembling a "real" hackathon \\
    Q5  & I prefer a hackathon format (short time and synchronous work) rather than a conventional project format (with more time and more asynchronous work)\\
    Q6  & Doing a project under a hackathon format was fun\\
    Q7  & The time (Thursday-Monday) was appropriate for the type of project we did\\
    Q8  & I was satisfied with the domain (e.g., urban mobility, COVID) chosen by my team\\
    Q9  & The chosen domain can influence the team's productivity in the project\\
    Q10  & I would participate in another course using a hackathon as a learning tool\\
    \multicolumn{2}{l}{\textbf{Working remotely}}\\
    Q11 & The use of the Discord tool was fundamental for communication with the teacher and mentors\\
    Q12 & I felt uncomfortable using a leisure tool (Discord) for an academic activity\\
    Q13 & The use of Discord was fundamental for the smooth functioning of teamwork\\
    Q14 & My team worked synchronously, supported by the Discord tool\\
    Q15 & My team constantly used Discord's voice channels\\
    Q16 & My team constantly used Discord's screen sharing \\
    Q17 & My team constantly used pair programming (shared screen or VSCode)\\
    Q18 & Pair programming was important for my team\\
  \bottomrule
\end{tabular}
\end{table}

\noindent\textbf{Participant Observation.} The observations were done by the teacher and two mentors (the third student mentor was not involved in the observation). Our participation consisted in being online on Discord during both types of follow up sessions (ad-hoc mentoring and scheduled checkpoints) and jump in the groups' voice channels and interact with them through voice and screen sharing (e.g., discussing a diagram, or as mere observers of who was online. 
We acted as Observer-as-Participant and Participant-as-Observe-- under Spradley's \cite{spradley} perspective would be equivalent to moderate and active participation, respectively -- 
, playing more than one role at different times and situations \cite{johnson2019educational}. 
During and after follow up sessions, condensed field notes \cite{spradley} were taken using text editor software. While ad-hoc mentoring consisted on being pro-active and asking if students needed any help, the checkpoints were more structured and followed a general script asking about group's accomplishments so far, main difficulties, new topics learned, sync or async group work, the next planned steps, etc, which helped to identify eventual problems. In both mentoring and checkpoint sessions, mentors and teacher took observation notes of other perceivable actions going on (e.g., eventual interactions among teams; which groups were using the voice channels or sharing screens at the moment). The generated notes were used as a data source. We also used the Discord logs of the text channels as a complement to those observations.

\noindent\textbf{Postmortem.} Reflection has an important role in approaches such as experiential learning and active learning, especially in a collective scenario \cite{raelin2001public,rodgers2002defining}. Students take a step back and reflect on their learning process through a different angle. In the week after the hackathon we conducted a project \textit{postmortem}~\cite{birk2002postmortem}, which is a consolidated practice in agile methods~\cite{reichlmayr2003agile} for reflecting about projects~\cite{andriyani2017reflection}. We gave them prompts asking what worked or did not work in: (1) the hackathon and course format; and (2)  their hackathon project itself, with an additional question of what were most important topics and technologies they learned during the project. 
Our goals were to have feedback on the intervention and to promote reflective thinking about their learning. 
The postmortem session took around 60 minutes, in the \#auditorium voice channel on Discord. The first 20 minutes consisted on 10 minutes for instructions and 5-10 minutes for group discussions in parallel. In the remaining 40 minutes, when we started recording the audio\footnote{An Informed Consent Form was signed by students clarified aspects regarding study goals and confidentiality on the collected information and the participants' identities.}, each group shared their reflections collectively to all other students. We transcribed the audio and analyzed it through a qualitative procedure given the nature of the information. The analysis followed two steps of the coding process we borrowed from Grounded Theory \cite{corbin2014basics}: open and axial coding. To help to structure the data, we used a shared spreadsheet tool and opted to use an inductive approach of coding, with codes generated from the transcription. To increase trustworthiness during the coding and further interpretation, the process was conducted by two researchers, each one coding the postmortem session, and, thereafter, comparing, discussing, and interpreting the results.

\section{Results}
\subsection{Survey Data}
We had 17 survey answers (just one student did not answer). Only 4 respondents (23.5\%) had not participated in a hackathon before. All remaining respondents had experience in in-person events, while one of them also had participated in online hackathons before. 

\begin{figure}[!ht]
\centering
\includegraphics[width=0.75\textwidth]{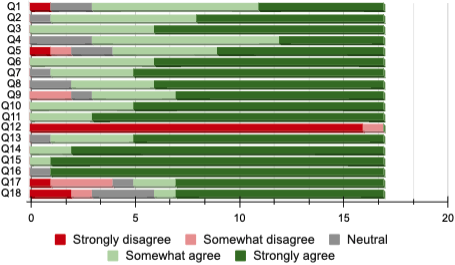}
\caption{Answers to Likert-scale questions of the survey.}
\label{fig:likert}
\end{figure}

\noindent\textbf{Motivation.} 
Q1 asked about excitement \textit{before} the hackathon. Most answers are an agreement (82\%), while 2 (12\%) were neutral and 1 (6\%) disagreed. In contrast, responses of Q2 and Q3 were virtually unanimous on being motivated (94\%) and engaged (100\%) during the hackathon, respectively. Just one response was neutral for Q2. As illustrated on \autoref{tab:questions}, to avoid ambiguity on Q2 and Q3 both sentences had a brief explanation we gave based on literature\cite{martin2017motivation}.

\noindent\textbf{Hackathon as a project format.} 
Q4 asked about the hackathon being authentic (i.e., a ``real'' hackathon), with 100\% agreeing. While 13 (76\%) respondents prefer the format of a hackathon instead of a conventional asynchronous project (Q5), 2 disagreed (12\%) and 2 were neutral (12\%). All respondents agreed that doing the project as a hackathon was fun (Q6). Concerning the duration (Thursday-Monday) of the hackathon (Q7), there was no disagreement and only one neutral response (6\%). 15 respondents (88\%) agreed they were satisfied with the domain chosen by their group (Q8), while the 2 (12\%) remaining respondents were neutral. 14 people (82\%) agreed that such choice influences productivity (Q9),  while 2 (12\%) disagreed and 1 (6\%) was neutral. All respondents agreed they would join another course using hackathons as a learning tool (Q10). 

\noindent\textbf{Remote work.} This block also had two general questions on using Discord. One asked about having using Discord before: 12 people (71\%) used it for gaming and as a communication tool at work (e.g., internships) and 5 (29\%) used it just for gaming. The other asked about using Discord in another course: 10 people (59\%) answered yes. For the remaining questions in the form of statements, all respondents agreed on the importance of Discord for communication with teacher and mentors (Q11). None of them felt bothered for using a leisure tool (Discord) in academic activities (Q12). Most respondents agreed (94\%) on the importance of Discord for teamwork (Q13), with only one neutral answer (6\%). All of them agreed their teams worked synchronously with the help of Discord (Q14), and voice channels were used constantly (Q15). A majority of respondents (94\%) agreed with the statement indicating they constantly used the screen sharing feature of Discord (Q16), while only one respondent was neutral (6\%). Pair programming being performed either through screen sharing or Visual Studio Code\footnote{The code editor used in all teams, with some groups using the Live Share extension.} (Q17) had an agreement in 12 answers (71\%), 1 neutral (6\%), and 4 (24\%) disagreements. On the importance of pair programming (Q18), 11 people (65\%) agreed with it, 3 (17.5\%) were neutral, and 3 disagreed. 

\begin{figure}[!ht]
\centering
\includegraphics[width=0.75\textwidth]{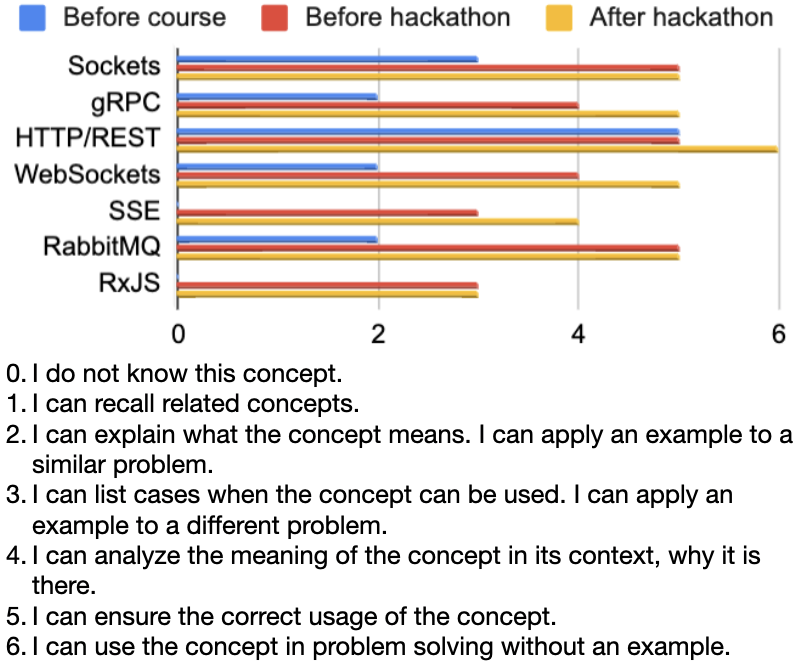}
\caption{Student's perception of learning on each course topic, under a perspective based on Bloom's Taxonomy.}
\label{fig:bloom}
\end{figure}

\noindent\textbf{Learning.} The median of responses about the students' perception regarding their learning (\autoref{fig:bloom}) indicate that they improved their knowledge on all topics during the course, except for HTTP/REST, with certain topics (Server-sent Events and RxJS) being unknown to most of them. The hackathon was fundamental to improve learning on gRPC, HTTP/REST, WebSockets, and Server-sent Events, whereas the remaining topics remained the same as before the hackathon (note that sockets were not explicitly used in any project).

\noindent\textbf{Open-ended questions.} The first open-ended question regarding additional comments on the usage of Discord generated some codes that we categorized as four positive perceptions: Satisfaction, Channels organization, Collaboration, and Mutual awareness of teams working. 
The only negative perception concerned our recommendation on using Discord as the official communication channel: \textit{``I didn't like the matter of `forcing' the use of text channels. It is not practical (discord notifications are bad) and gives a `big brother' feeling.''} The generated categories regarding the analysis of the overall positive aspects of their experience in doing a project during a hackathon were: Putting knowledge into practice, Real case scenario, Collaboration, Mentoring, and Optimized time management. Their feedback on negative aspects led our analysis to the following categories: Student schedule availability, time constraints, decision of project scope, limited coverage of topics in the project.

\subsection{Participant Observation}
On the first day (Thursday) students focused on the choice of the data set and the definition of their challenge. On Friday they discussed architecture, testing technologies, and APIs. Most of the development focused on the solution that happened during the weekend. On Monday the projects were almost complete, and students focused on adjustments or minor bug fixes.


\noindent\textbf{Discord usage.}
All participants were already familiar with the communication tool. The common behavior in groups was every team member connected to the voice channel as if they were sharing the same physical space. They could be silent for some moments, they could be talking, they could leave or join at any time. A good metaphor for that would be a shared room or office. Students could also start a transmission to share their screen so they could do an activity together (e.g., coding, debugging, designing, or discussing the architecture). In the mentoring sessions, students discussed various issues (e.g., bugs, architectural decisions, questions about technology usage) over the voice channel, and when needed, they would share their screens to show more details. During checkpoints, the interaction was frequently voice-only, but sometimes they would share their screen to show something they needed help with.


\noindent\textbf{Text channels as support to voice channels.} The text channels of the groups were used as support to voice conversations for sharing links, architecture drawings or code snippets. The technology channels were not used at all, except for the \#ui channel, where some groups shared information and tips. 
A student mentioned not using them because he already had an idea about the technologies seen in the semester and knew where to find resources, if necessary, while UI (User Interface) was not part of the course.

\noindent\textbf{Learning.} During checkpoints students said they were not necessarily learning new things but rather reinforcing what they had already seen. They also said it was an opportunity to go deeper in topics they missed assignments or did not know much. One student mentioned that \textit{"The biggest difference from the assignments is that as we now have to use more technologies, we sometimes spent more time discussing than implementing. The assignments were self-contained. Now we had to discuss the use case and which technology was better.}"

\noindent\textbf{Combination of synchronous and asynchronous work.}
Groups worked mixing asynchronous and synchronous work in different proportions. Since many of them had a day job or internship, most of their work in the hackathon was conducted at night or over the weekend. During the day they worked mostly asynchronously, while at night they would all be online and most of the times sharing their screen and doing pair programming. The two groups most frequently observed working synchronously had the highest hackathon evaluation and grades. The two groups that favored asynchronous work had an average and a good grade, respectively. 


\noindent\textbf{Healthy habits.} During checkpoints participants typically mentioned their pauses for meals, resting and even game playing. As they were home, students mentioned working until late hours (also confirmed by a mentor observing 3 groups working at 1:30 AM), but also emphasized they were sleeping well. As a student said, he did sleep \textit{"... especially because I was working right next to my bed"}.

\noindent\textbf{Collaboration between teams.}
There were few interactions between groups about doubts and it was in groups in which the members already knew each other. These participants would eventually "visit" another team's voice channel.

\noindent\textbf{Friendly ambiance.} Although online, participants appeared to be at ease. The Groovy music bot (a Discord extension installed on the server after their request on day 1) also helped in this aspect. The  \#off-topic channel was relatively used for leisure, where participants shared memes, jokes and at some times invited others to play games.

\subsection{Postmortem}
The codes in this analysis were condensed into five categories below. 
The range S1-S9 represents the 9 students who spoke in the session. 

\noindent\textbf{Motivational Aspects}. This category was originated from codes such as engagement, productivity, fatigue, and comfort zone. We perceived a feeling of engagement and productivity among the teams, as quoted by S1: \textit{``What went right was the engagement and synergy of the team. The guys were really committed.''} Although expected in such an intensive event, only one team reported the fatigue issue due to the considerable time took to develop some features. Still, many students remained in their comfort zones resulting in some of them not experimenting with all the topics covered during classes; they argued that, given the time frame and the competition, they would rather use something they already knew to accomplish an on-time good result. Moreover, some also highlighted they already had the opportunity to touch all the topics on course assignments.

\noindent\textbf{Hackathon Format}. The usage of Discord was key to arrange students into text/voice channels per group and to facilitate screen sharing. 
Nevertheless, some students complained about missing Discord's notifications. Students ended up using other alternatives such as Telegram to overcome this issue. Another positive aspect was the freedom to choose the domain of projects which resulted in students searching for data sets closer to real projects, a motivator pointed by some of them: \textit{``I enjoyed [..] especially the choice of the project. I was afraid the hackathon was not going to be that flexible''} (S9). The format of a hackathon was preferred over a regular project assignment that would take weeks. But they also faced technical issues with some data sources (e.g., APIs returning wrong data, broken data sets). Some students mentioned schedule conflicts of the hackathon with final assignments from other courses.

\noindent\textbf{Management.} Students were able to manage their tasks and time efficiently; the only issue was that some of the students felt blocked by other students tasks, while others reported it running smoothly, such as S5: \textit{``we distributed responsibilities in such a way that everyone could work asynchronously without blocking each other.''} Group members were already familiar or even friends among themselves, resulting resulted in a natural team-building process and helping on activities like pair programming: \textit{``we already knew each other [..] it was easier to use the voice channel and share the screen''} (S2).

\noindent\textbf{Learning and Practice.} All groups reported being able to apply their classes knowledge with no problem: \textit{``We could excel in the hackathon because of the base provided by the course assignments''} (S2). They also enjoyed the hackathon experience because their projects were similar to real scenarios:
\textit{``[...] real-world technologies that companies use. The type of thing that you don't say out of the blue `I want to learn RabbitMQ by myself'''} (S8). But students were confused about the evaluation process, not knowing if they had to focus on the technologies or the presentation since there was a reward. Even after they were told their grades would not depend on the prize, it was still an issue stated by most students. They also gave two suggestions: (1) implementing asynchronous classes, similar to an inverted classroom \cite{lage2000inverting} and (2) Discord being used early in the semester so students could interact earlier.

\noindent\textbf{Communication.} The groups were in constant interaction and they mentioned that, thanks to Discord's ability to have multiple discussion channels, mentors could ask each group from time to time if they needed any help, and groups could also help each other. Nonetheless, there was one student (S4) that was not pleased by the communication tool enforcement: \textit{``Discord is a good platform, but it would be better if we had the option to choose.''} 

\section{Lessons Learned}
After converging the analysis of our data sources, we highlight lessons learned for those intending to apply similar interventions.

\noindent\textbf{Discord as a communication tool.} The tool and the way the channels were configured were fundamental for the successful participation and involvement of students. Many students already use it for gaming, therefore this familiarity would make the adoption of a classroom Discord server a seamless transition. 

\noindent\textbf{Online radical collocation.} The radical collocation that helps achieve better results in  hackathons~\cite{trainer2016hackathon} could be adapted to the immersive online environment provided by Discord. The voice channels and screen sharing allowed to do things like pair programming remotely. It all made them feel closer and have a sense of virtual (i.e., online) collocation. Teams that spent more time doing online synchronous work had better projects and results. 

\noindent\textbf{Socialization boosts engagement.} Socialization is a key motivator in a learning environment. Besides the synchronous collaboration among team members toward a common goal, participants interacted to help each other and to play games. A Discord server can host simultaneous voice channels and other participants can see who is connected to a channel. This mutual awareness of teams working in parallel was reported as something stimulating. 

\noindent\textbf{Consolidation of Learning.} The practice simulating a realistic scenario was appreciated. There was not much time to learn new things and students remained in their comfort zones, but consolidating topics they knew. The simple self-assessment survey question indicated improvement in many topics. The lack of task rotation among team members limits the number of topics practiced by each student~\cite{gama2019developing}, but they said it would also happen in a regular project.

\noindent\textbf{Duration.} Four days seemed to be an appropriate duration, with teams having time to organize their tasks and work without any major rush. They valued eating and sleeping well, in contrast to the sleep deprivation scenario that is typical from hackathons \cite{uys2019hackathons}.

We also enumerate some drawbacks concerning aspects that we will change when if running this activity again in a remote setting: 

\noindent\textbf{Evaluation criteria.} Having two evaluations -- one to announce a symbolic hackathon winner and later one as a detailed evaluation for grading -- with slightly different criteria caused some confusion. They have to be clear so students know which criteria to prioritize. 

\noindent\textbf{Softer enforcement of communication channel.} 
Students indeed enjoy using Discord and even suggested using it earlier in the semester. However, enforcing it as the only communication channel for the group generated discomfort and a feeling of being under surveillance. Other tools would be more natural and effective while they are not connected to the server. A similar feeling of hindered communication due to making the usage of Discord mandatory has also been reported in the context of educational game jams \cite{gledhill2019game}.


\section{Conclusions}

With the sudden change to emergency remote teaching, students have faced a lack of engagement in remote learning activities for many reasons, especially the reduced socialization. In this experience report, we detail our intervention of using an online hackathon as an educational tool to engage students in group work.  We provided details on the activity organization, and reported an analysis of the mentors' observations and feedback collected from students. One of the most important aspects we can highlight is the usage of Discord as a communication tool. Its configuration and the dynamic we conducted helped to create an intense collaborative experience. Students could socialize online and synchronously interact with each other, having the sense of being virtually collocated. This socialization made possible using this environment and the challenge-oriented nature of the project was important to engage students who reported excitement and motivation. They considered the duration (\~96h) as adequate, with time to rest while practicing and consolidating knowledge in a realistic project without the rush of most hackathons (12 to 48 hours). They perceived the event helped them to enhance their knowledge in some of the course topics.

Some drawbacks are the possible biases in the analyses, such as participant bias to please the teacher, although students were stimulated to criticize and appeared to be at ease, or researcher bias, which were minimized with the participation of two other researchers. There is also a lack of gender perspective, as no women participated. But we hope to address this issue in another semester. 

\bibliographystyle{unsrt}
\bibliography{references}

\end{document}